\documentclass[seceq]{ptptex}

\usepackage{graphicx}
\newcommand{\del}{\partial}

\title{
The influence of fluctuations on thermodynamics  near chiral phase transition%
}

\author{
Krzysztof \textsc{Redlich}$^{1,2}$, Bengt \textsc{Friman}$^{3}$, and  Vladimir \textsc{Skokov}$^{3}$%
}

\inst{%
$^{1}$
Institute of Theoretical Physics, University of Wroclaw, PL--50204 Wroc\l aw, Poland
\\
$^{2}$
Theory Division, CERN, CH-1211 Geneva 23, Switzerland
\\
$^{3}$
GSI Helmholtzzentrum f\"ur Schwerionenforschung, D-64291
Darmstadt, Germany
}

\abst{

We discuss the influence of 
fluctuations on thermodynamics  near the chiral phase transition within
Polyakov loop extended
 quark--meson model based on the  functional renormalization group (FRG) method. We include the gluon fields in the FRG flow equation self-consistently on the mean-field level.
 We focus on  the properties  of the phase diagram and   net-baryon number fluctuations.

}

\begin{document}

\maketitle

\section{Introduction}
 \label{sec:int}
Understanding the phase structure and critical properties of strongly
interacting matter is one of the central problems addressed in studies of QCD
thermodynamics. In  the context of heavy ion experiments of particular importance is to find measurable observables that can be used as   probes of 
the  deconfinement and  chiral phase transitions expected in QCD \cite{fk}.
Effective chiral models  were shown to be very useful to study critical phenomena in the strongly interacting system and to quantify  the phase transition due to deconfinement and chiral symmetry breaking\cite{hmm}.

 Particularly interesting are models like the Polyakov loop extended
Nambu--Jona--Lasinio (PNJL) \cite{PNJL} or quark--meson (PQM) \cite{PQM} models  which can account for both deconfinement  and chiral symmetry breaking.
Both 
 models can reproduce essential
properties of  QCD
thermodynamics obtained in the Lattice Gauge Theory    already within  the mean-field (MF)
approximation. However, to
correctly account for  critical behavior  and scaling properties  near the
chiral phase transitions
one needs to go beyond the mean-field approximation and include  fluctuations and
non-perturbative dynamics. This can be achieved  by using methods based on the functional renormalization group
(FRG)~\cite{Berges:review,our,org,orgl,s}.

 In this paper we summarized  the influence  of non-perturbative  effects  in the presence of a gluonic background  on the 
 thermodynamics the   near chiral phase transition based on the PQM model. We compare the phase diagram and  properties of net-quark number fluctuations obtained within the FRG and MF calculations. Such a comparison  quantifies explicitly the  influence  of
 mesonic fluctuations near the chiral phase transition at finite temperature and density\cite{org,orgl}.

\begin{figure}\hspace*{0.5cm}
\includegraphics*[width=6.3cm]{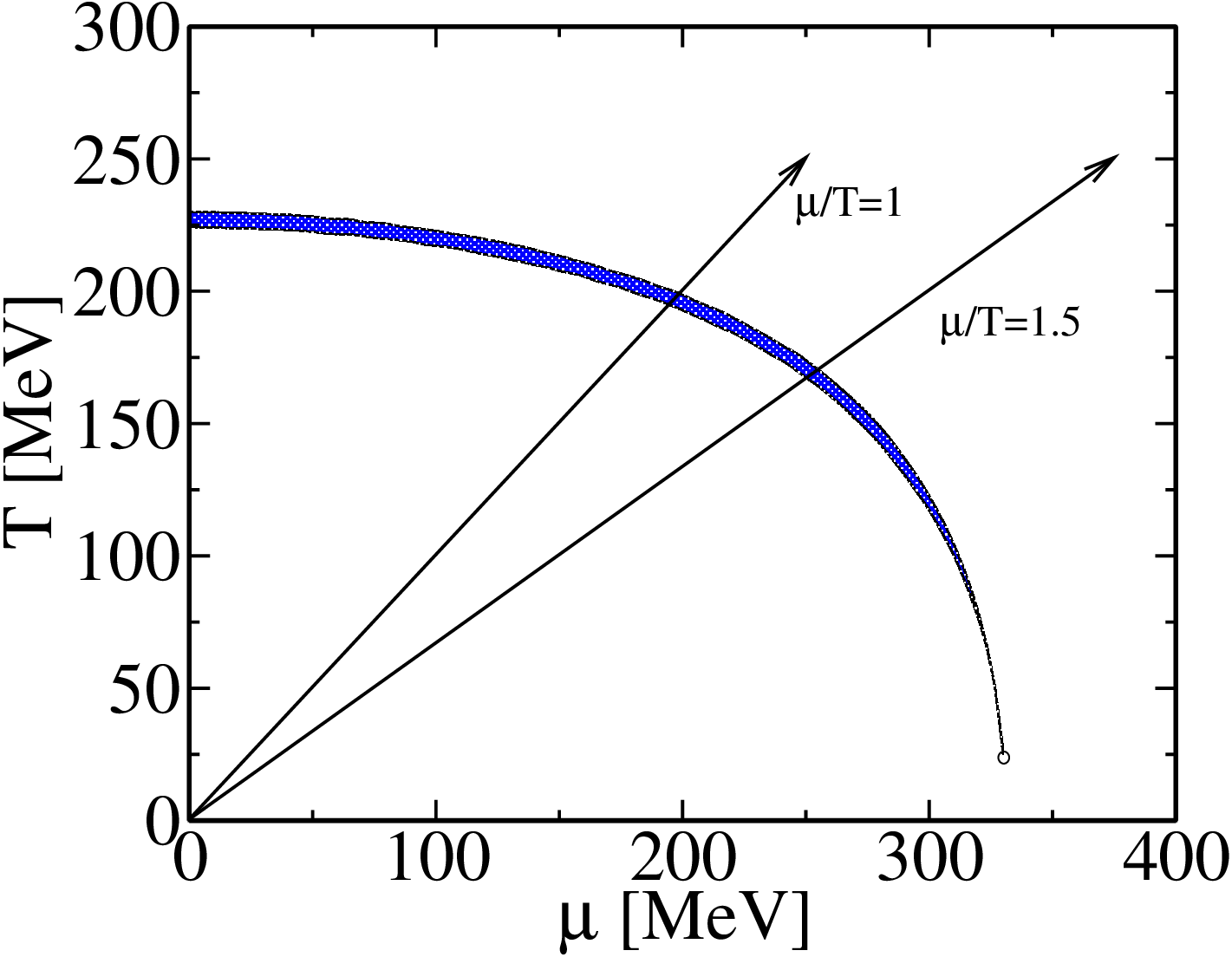}
\includegraphics*[width=6.3cm]{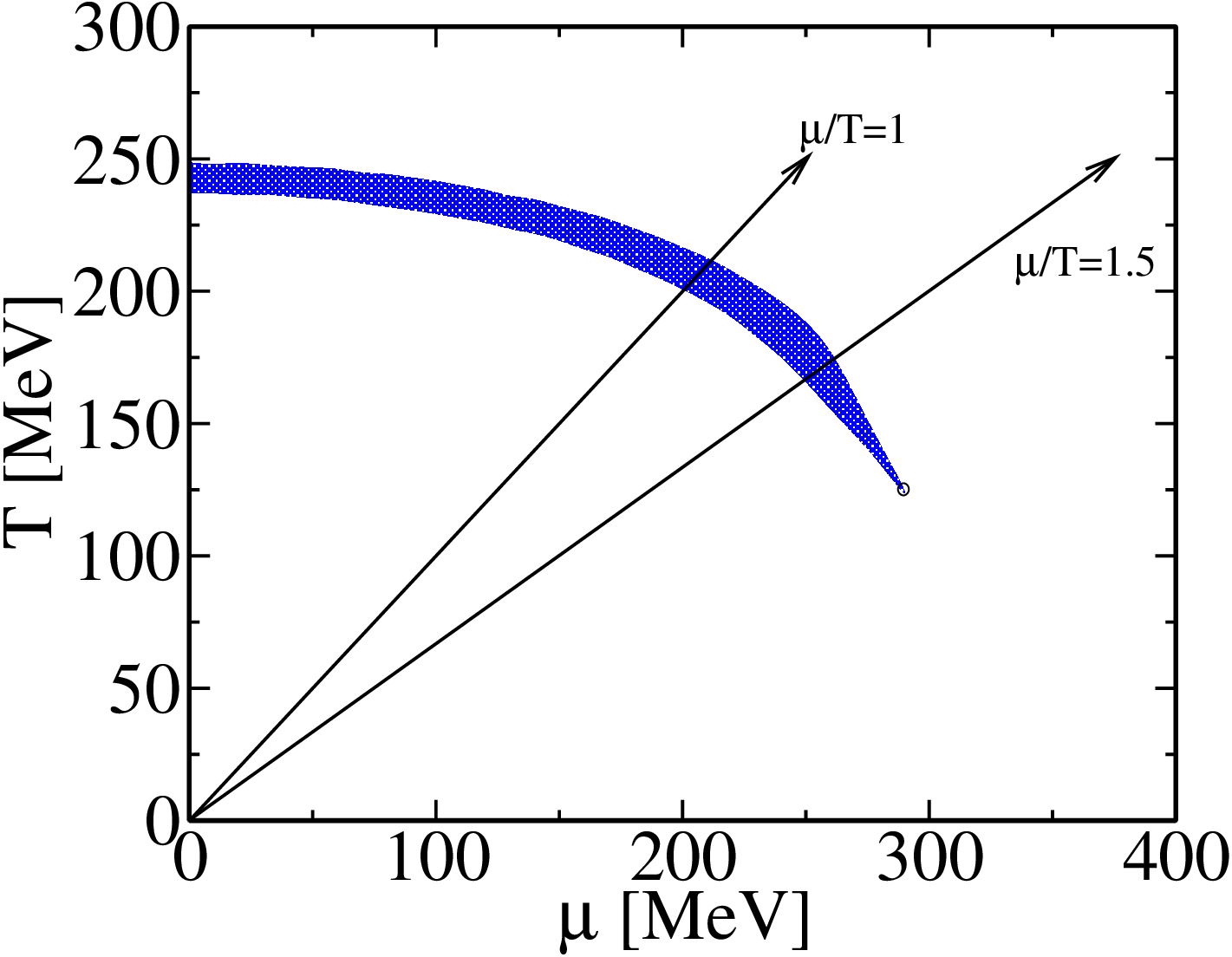}
\caption
{
The phase diagrams for the  PQM  model in the mean-field approximation (left panel)
and  in the  functional renormalization group approach (right panel).
The arrows indicate   lines corresponding  to different values of $\mu/T$. The shaded lines  define  a critical region (see text).
}
\label{fig:PD}
\end{figure}

\section{The FRG flow equation in a background  gluon field}
\label{sec:znf}
The role of   non-perturbative effects in the thermodynamic description  of   the PQM  model
can be studied  within the functional renormalization group (FRG) method.
The FRG  is based on an infrared  regularization with the momentum
scale parameter  of
the full propagator which turns the corresponding effective action into the  scale $k$-dependent functional
$\Gamma_k$~\cite{Berges:review}.
In general, in the PQM model, the formulation of the  FRG flow equation
requires an
implementation of the Polyakov loop as a dynamical field. However, in the current  formulation we treat the Polyakov loop as a background field which is introduced self-consistently on the mean-field level \cite{org}.
Following our previous work~\cite{org,orgl} we formulate the
flow equation for the scale-dependent grand canonical potential as
\begin{eqnarray}\label{eq:frg_flow}
\del_k \Omega_k(\ell, \ell^*; T,\mu)&=&\frac{k^4}{12\pi^2}
 \left\{ \frac{3}{E_\pi} \Bigg[ 1+2n_B(E_\pi;T)\Bigg]
 +\frac{1}{E_\sigma} \Bigg[ 1+2n_B(E_\sigma;T)
    \Bigg]   \right. \\ \nonumber && \left. -\frac{4 N_c N_f}{E_q} \Bigg[ 1-
N(\ell,\ell^*;T,\mu)-\bar{N}(\ell,\ell^*;T,\mu)\Bigg] \right\}.
\end{eqnarray}
Here $n_B(E_{\pi,\sigma};T)$ is the bosonic  distribution function
\begin{equation*}
n_B(E_{\pi,\sigma};T)=\frac{1}{\exp({E_{\pi,\sigma}/T})-1}
\end{equation*}
with the pion, $E_\pi = \sqrt{k^2+\overline{\Omega}^{\,\prime}_k}$,  and sigma, $E_\sigma
=\sqrt{k^2+\overline{\Omega}^{\,\prime}_k+2\rho\,\overline{\Omega}^{\,
\prime\prime} _k}$
energies,
where the primes denote derivatives with respect to $\rho=(\sigma^2+{\vec\pi}^2)/2 $ and
$\overline{\Omega}=\Omega+c\sigma$ with $c=m_\pi^2 f_\pi$ being the  external symmetry breaking term.
The functions $N(\ell,\ell^*;T,\mu)$ and $\bar{N}(\ell,\ell^*;T,\mu)$ which are  defined
by
\begin{eqnarray}\label{n1}
N(\ell,\ell^*;T,\mu)&=&\frac{1+2\ell^*\exp[\beta(E_q-\mu)]+\ell \exp[2\beta(E_q-\mu)]}{1+3\ell \exp[2\beta(E_q-\mu)]+
3\ell^*\exp[\beta(E_q-\mu)]+\exp[3\beta(E_q-\mu)]},\nonumber  \\
\bar{N}(\ell,\ell^*;T,\mu)&=&N(\ell^*,\ell;T,-\mu),
\label{n2}
\end{eqnarray}
are fermionic distributions which are modified because of  coupling to gluons. The
$
E_q =\sqrt{k^2+2g^2\rho}
$
is the quark energy.
The minimum of the thermodynamic potential  is determined by the stationarity
condition
\begin{equation}
\left. \frac{d \Omega_k}{ d \sigma} \right|_{\sigma=\sigma_k}=\left. \frac{d
\overline{\Omega}_k}{ d \sigma} \right|_{\sigma=\sigma_k} - c =0.
\label{eom_sigma}
\end{equation}

The flow equation~(\ref{eq:frg_flow}) is solved
numerically with the initial cutoff $\Lambda=1.2$ GeV (
see details in Ref.~\cite{org,orgl}).
The initial
conditions for the flow  are chosen
to  reproduce 
the physical pion mass $m_{\pi}=138$ MeV, the pion decay constant
$f_{\pi}=93$ MeV, the sigma mass $m_{\sigma}=600$ MeV and the constituent quark mass $m_q=300$ MeV
at the scale $k=0$ and for zero temperature and chemical potential $T=\mu=0$.

By solving  equation  (\ref{eq:frg_flow})  one obtains the thermodynamic potential
for   quarks and mesons, $\Omega_{k\to0} (\ell, \ell^*;T, \mu)$,
as a function of  Polyakov loop variables, $\ell$ and $\ell^*$. The  full thermodynamic potential $\Omega(\ell, \ell^*;T, \mu)$ in the PQM model which
includes quarks, mesons and
gluons degrees of freedom
is obtained by adding  to  $\Omega_{k\to0} (\ell, \ell^*;T, \mu)$ the
effective gluon  potential, ${{\cal U}(\ell,\ell^{*})}/{T^4}=
-{b_2(T)}\ell^{*}\ell
-{b_3}(\ell^3 + \ell^{*3})
+{b_4}(\ell^{*}\ell)^2
$. Thus,
\begin{equation}
\Omega(\ell, \ell^*;T, \mu) = \Omega_{k\to0} (\ell, \ell^*;T, \mu) + {\cal U}(\ell, \ell^*),
\label{omega_final}
\end{equation}
with
 parameters $b_i$
chosen  to reproduce the equation of state of the  pure SU$_c$(3) lattice gauge theory \cite{PNJL}.

At a given temperature and chemical potential,
the Polyakov loop variables, $\ell$ and $\ell^*$, are determined by the
stationarity conditions:
\begin{eqnarray}
\label{eom_for_PL_l}
\frac{ \partial   }{\partial \ell} \Omega(\ell, \ell^*;T, \mu)  =0, ~~
\frac{ \partial   }{\partial \ell^*}  \Omega(\ell, \ell^*;T, \mu)   =0.
\label{eom_for_PL_ls}
\end{eqnarray}
The thermodynamic potential  (\ref{omega_final}) does not contain   contributions of  statistical modes with momenta
larger than  the cutoff $\Lambda$.
In order to obtain the correct high-temperature behavior of
  thermodynamics   we  need to supplement the
FRG potential with the  contribution of  high-momentum states.
For $k > \Lambda$ we use
the flow equation for quarks interacting with the  Polyakov loops \cite{org},
\begin{eqnarray}\label{eq:qcdflow}
\del_k \Omega_k^{\Lambda}(T,\mu)&=&-\frac{N_c N_f k^3}{3\pi^2}
 \Big[ 1-
N(\ell,\ell^*;T,\mu)-\bar{N}(\ell,\ell^*;T,\mu)\Big]
\end{eqnarray}
where the dynamical quark mass is neglected.

To obtain the complete thermodynamic potential of the PQM model   we   integrate
Eq.~(\ref{eq:qcdflow})  from $k=\infty$ to
$k=\Lambda$ where we switch to the PQM flow
equation (\ref{eq:frg_flow}).

\begin{figure}\hspace*{0.5cm}
\includegraphics*[width=6.4cm]{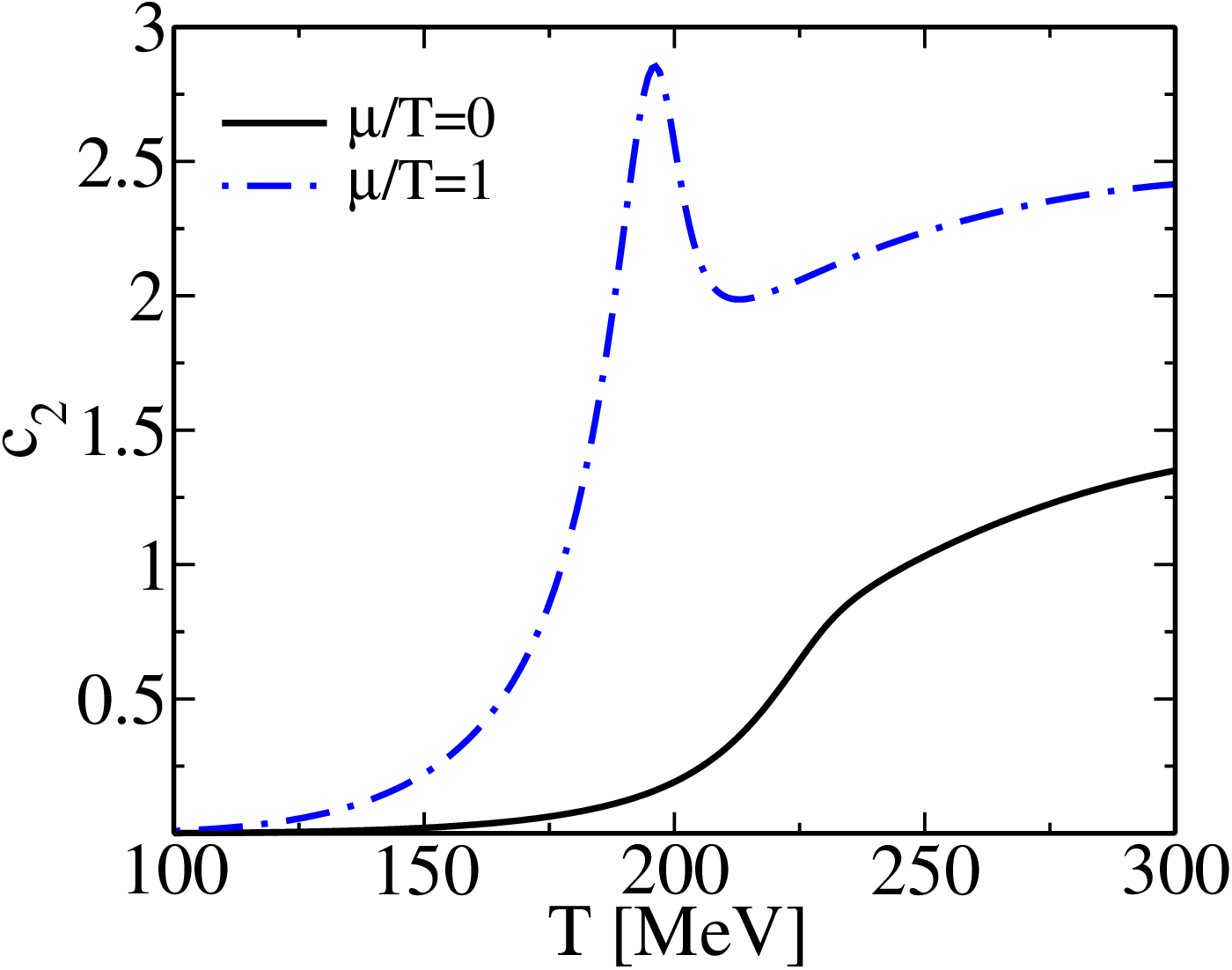}
\includegraphics*[width=6.1cm]{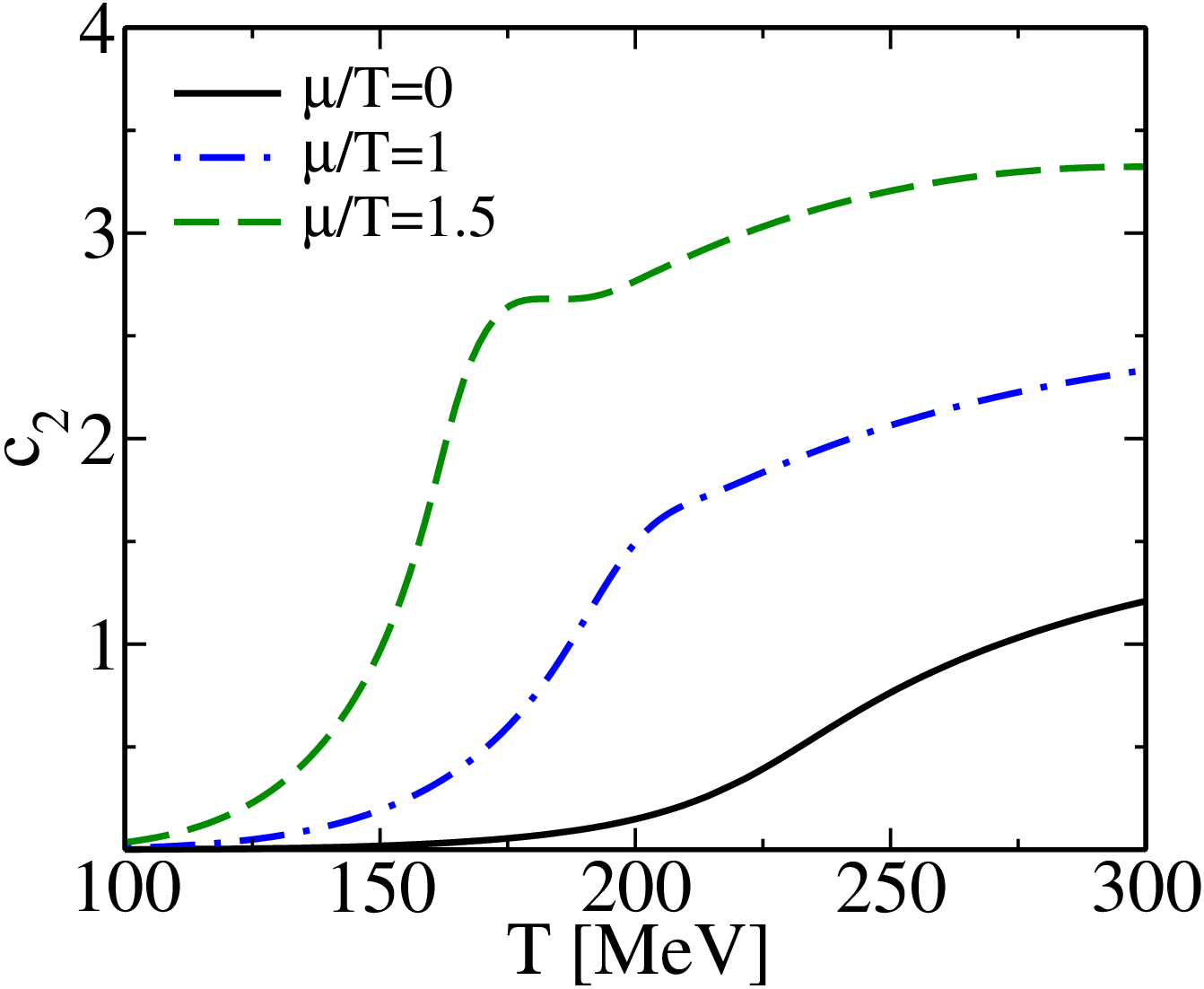}
\caption{
The coefficient $c_2$ as a function of temperature  for different values of $\mu/T$
for the {  PQM}  model
in the mean-field approximation (left panel) and
in the FRG approach (right panel).  }
\label{fig:c2}
\end{figure}
\begin{figure}\hspace*{0.5cm}
\includegraphics*[width=6.4cm]{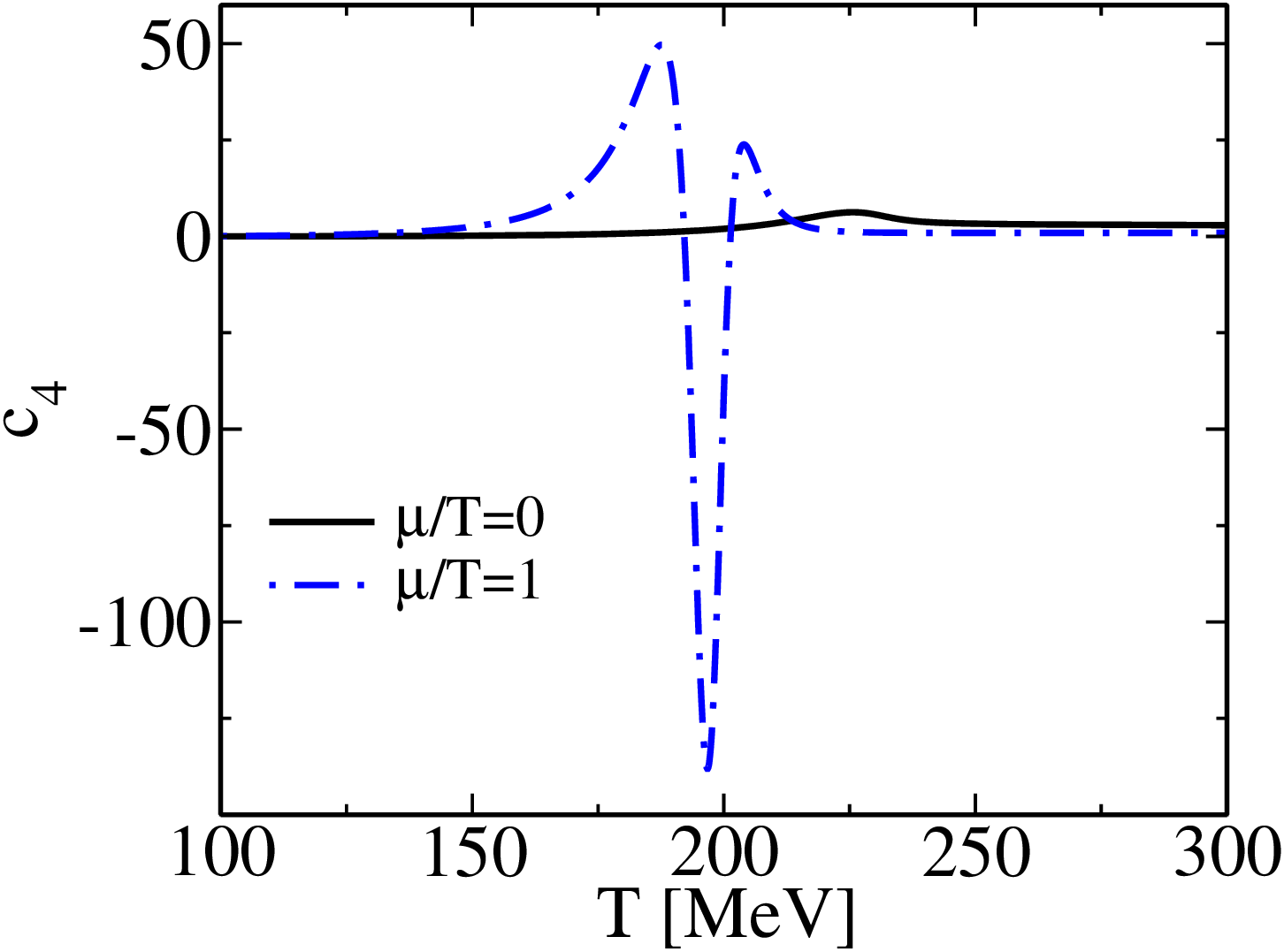}
\includegraphics*[width=6.1cm]{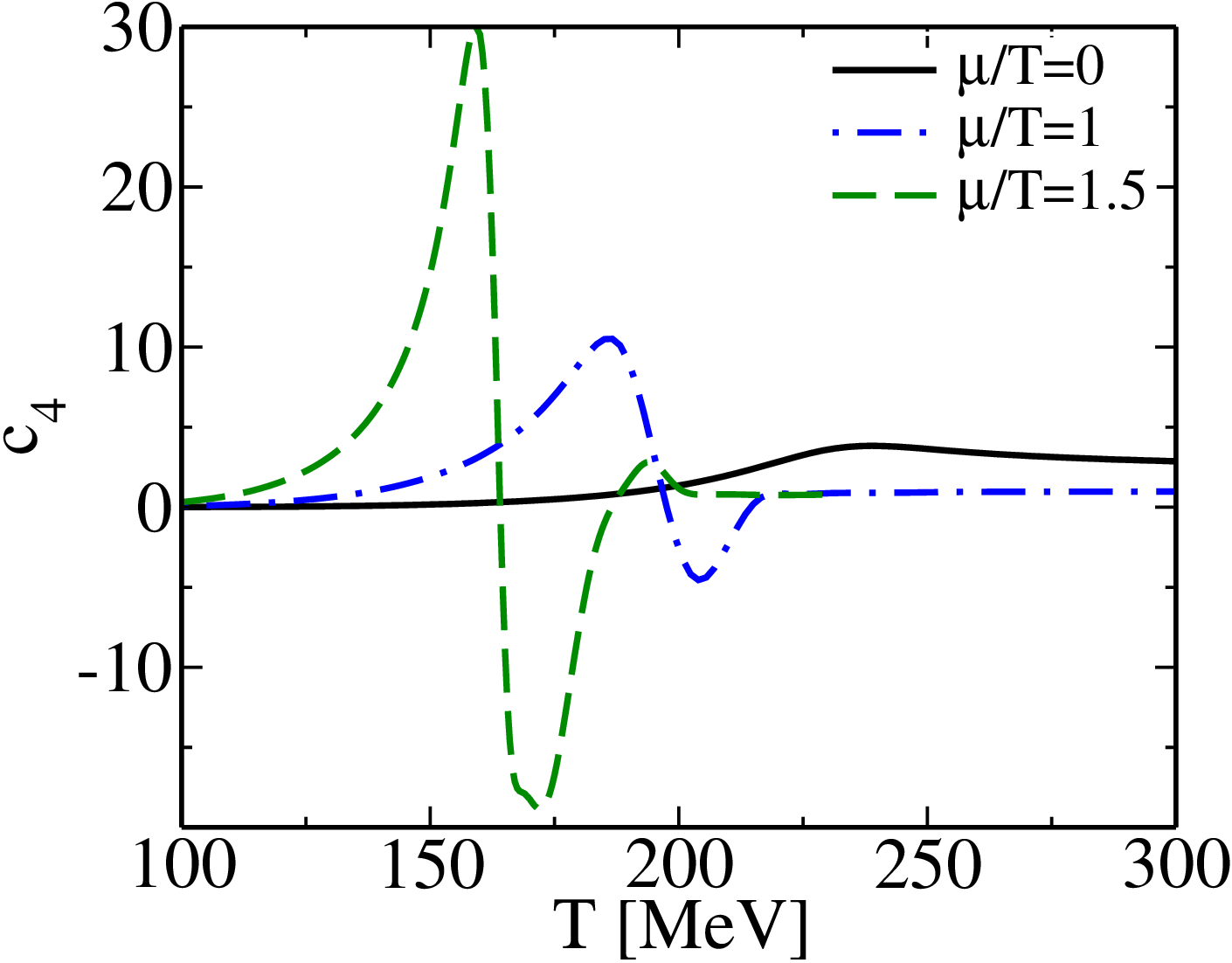}
\caption{
The coefficient $c_4$ as a function of temperature for different values of $\mu/T$
for the {  PQM} model in the mean-field approximation (left panel) and
in the FRG approach (right panel).
}
\label{fig:c4}
\end{figure}
\begin{figure}
\includegraphics*[width=6.5cm]{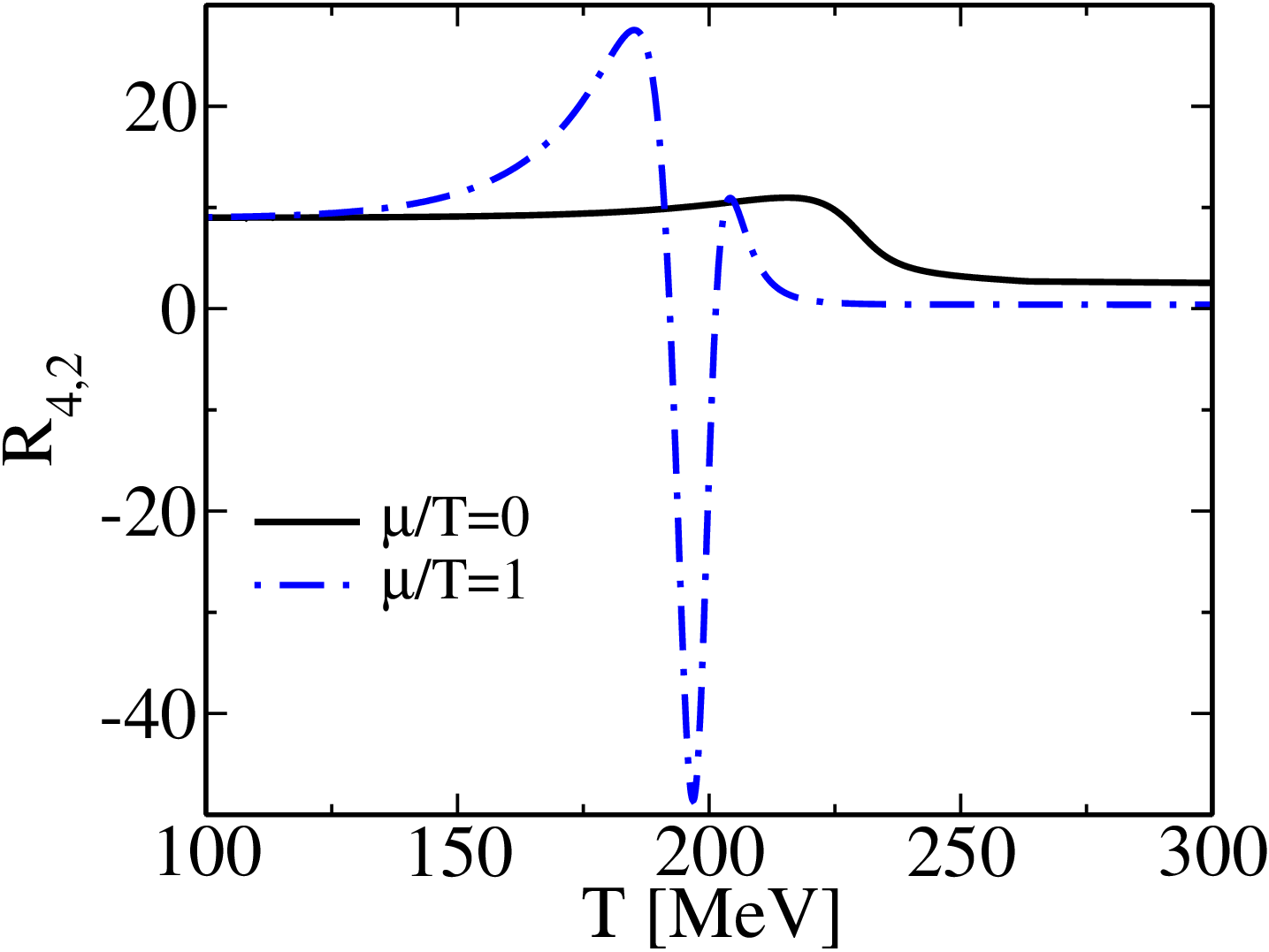}
\includegraphics*[width=6.3cm]{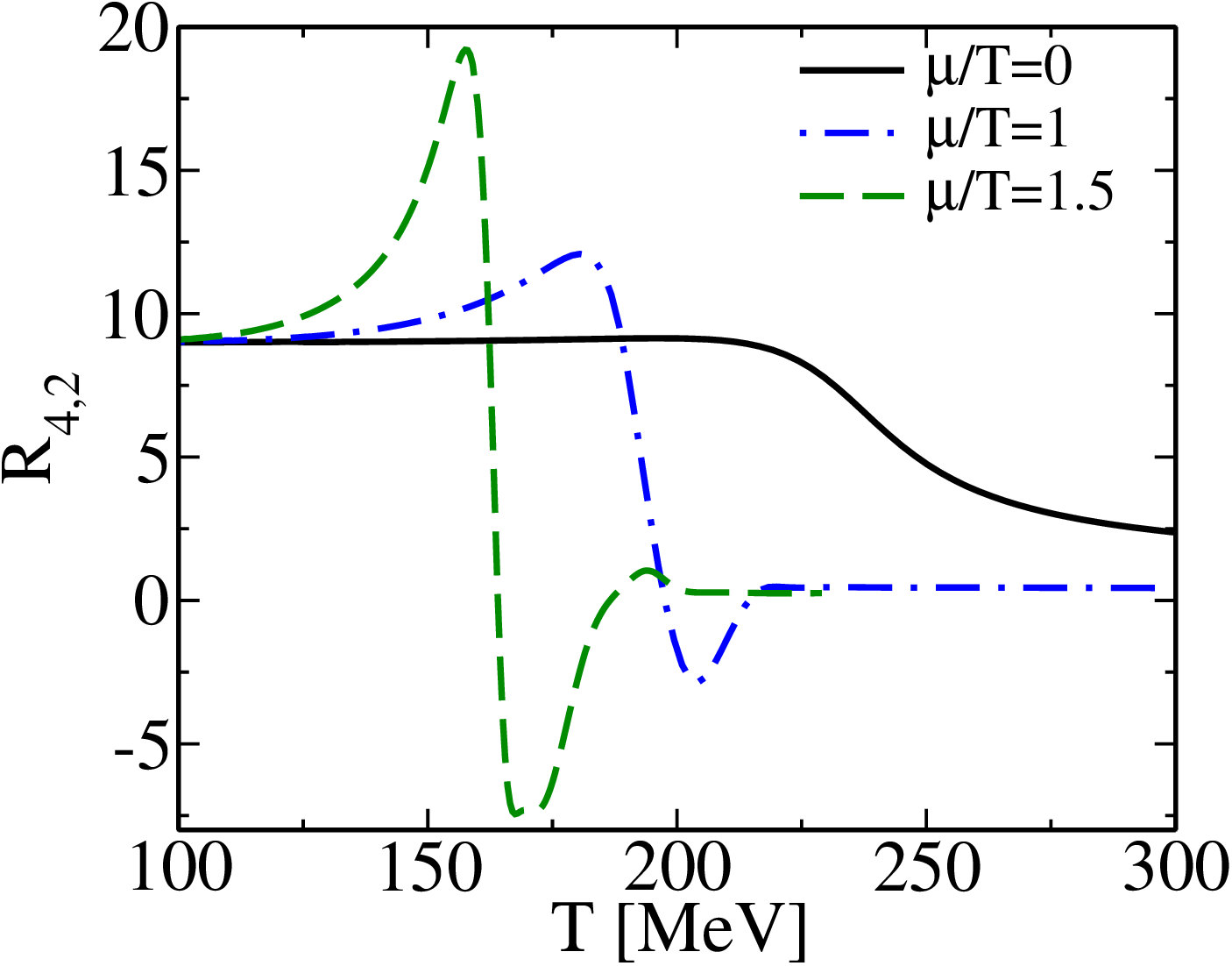}
\caption{
The kurtosis $R_{4,2}$ as a function of temperature for different $\mu/T$
calculated  in  the { PQM} model  under  the mean-field approximation (left panel) and
in the FRG approach (right panel).
}
\label{fig:r}
\end{figure}

\section{Thermodynamics of the PQM model in the presence of quantum fluctuations and gluonic background}
\label{sec:sum}

In order to illustrate  the importance of mesonic fluctuations and gluonic background on thermodynamics in the PQM model
we will compare the FRG results  with those obtained under a mean-field  approximation. In the latter,
both quantum and thermal fluctuations are neglected and the mesonic fields
are replaced by their classical expectation values. The formulation of the thermodynamic potential in the MF approximation can be found in Refs. \cite{PQM,our}.

Fig.~\ref{fig:PD}  shows the phase
diagrams of the PQM model obtained in  the  FRG approach  and  in the  mean-field
approximation. For the  physical pion mass and
moderate values of  the chemical potential  the PGM model exhibits a smooth cross-over chiral transition.    In Fig. \ref{fig:PD} we  define the  transition region   as a band where
 the temperature derivative of the order parameter exhibits  $5\%$-deviations
from its maximal value. At larger $\mu$ the cross-over line terminates at the  CEP where the transition is   second-order and belongs to the  universality class of the three  dimensional  $Z(2)$-spin system.

Comparing the resulting phase diagrams of the PQM model obtained within MF  and FRG approach
we find  a clear shift of  the position of the chiral phase boundary to  higher
temperatures  due to mesonic
fluctuations.  Changes  in the phase diagram  were previously reported  in the
quark-meson  model within
the FRG approach~\cite{PQM}. However, here  due to  gluonic background,  which is explicitly included in our FRG calculations, we find  a significant shift of the  CEP to higher temperature \cite{orgl}.

The  fluctuations of conserved charges were shown to be an excellent probe of the chiral and deconfinement transition expected in strongly interacting medium \cite{fk,Ejiri}. A particular role  was  attributed to   net-baryon number  fluctuations and their higher moments which  are characterized by the generalized susceptibilities,
\begin{equation}
c_n(T)=\frac{\del^n[p\,(T,\mu)/T^4]}{\del(\mu/T)^n}.
\end{equation}
The temperature and   chemical potential dependence of  the net-quark number susceptibility $c_2$ in the PQM model is shown in Fig.~\ref{fig:c2}
for the MF  and the FRG calculations.  At vanishing $\mu$   the coefficient  $c_2$ increases monotonously with temperature, similar to the thermodynamic pressure.
However, at finite chemical potential,  the susceptibility $c_2$  develops a peak structure.  The amplitude of
this peak increases with chemical potential towards the CEP where $c_2$
diverges. In the high temperature phase the $c_2$  converges  to
the Stefan Boltzmann value,
$
c_{2}^{ SB} =2[ 1 + 2({3\mu}/{\pi T})^2]
$.

As it is seen in  Fig.~\ref{fig:c2}  the peak structure
in $c_2$  is more pronounced in the mean-field  at  $\mu/T=1$  than
in the  FRG  at  $\mu/T=1.5$. This is in spite of the fact that the  location of the   CEP in the FRG   is closer  to the  $\mu/T=1.5$   than
 the corresponding  one for the mean-field approximation at  $\mu/T=1$ (see Fig. \ref{fig:PD}).
This  shows that  the criticality  of  $c_2$  as a function of
a distance to the CEP appears earlier in the mean-field than in the FRG calculations.

The generalized susceptibility $c_2$ is  sensitive to  changes in  the  chemical potential and is clearly influenced    by   meson  fluctuations and  the gluon background. The $c_2$ is strictly positive for all values of $\mu$ and T.
At finite $\mu$  the positivity   is not preserved for $c_n$-moments  with $n>2$.

 Fig.  \ref{fig:c4} shows  the  fourth  order cumulant moment  of the net-quark number  for different values of  $\mu/T$. The fourth order cumulant is strictly positive for vanishing chemical potential. However,  for
higher values of $\mu/T$,  $c_4$ becomes negative in the  vicinity  to the  cross-over transition.  The  chemical potential independent  Stefan-Boltzmann limit
$
c_{4}^{ SB} ={2 N_cN_f}/{\pi^2}$
is  reproduced at  temperatures $T>>T_c$.
 Comparing the MF  with  the FRG results for $c_4$  it is clear that  meson fluctuations are essentially modifying properties of   quark  susceptibilities. In the transition region $c_4$  is suppressed in the FRG relative  to the MF results.
The meson fluctuations quantified by the FRG method
 provide smoothing of  net-quark susceptibilities  near the crossover transition which  is substantially   broader in the FRG than in the MF approach.

Discussing fluctuations of conserved charges, a  particular role is attributed to the so called  kurtosis of  net-quark number fluctuations which 
is given by the ratio,
$
 R_{4,2}={c_4}/{c_2}.
$~\cite{Ejiri}
This key observable is not only sensitive  to the  chiral but also to  the confinement-deconfinement
transition. In the asymptotic regime  of high and low temperature and  at vanishing chemical potential  the kurtosis
reflects the  quark content of the baryon-number carrying effective degrees of freedom~\cite{Ejiri}.
Therefore, at low $T$ in the confined  phase,  $R_{4,2}\simeq
N_q^2=9$ while in an ideal gas of quarks $R_{4,2}\simeq 1$. This property was shown to be independent from the particular form of the mass spectrum and the number of degrees of freedom in a system.

Fig. \ref{fig:r}
 shows the $c_4/c_2$ ratio    calculated  as a function of temperature along different paths in the $(T, \mu)-$plane quantified by  fixed $\mu/T$.
In the PQM model
the kurtosis shows an expected  drop from $R_{4,2}\simeq 9$ to $R_{4,2} \simeq 1$ in the transition
region. At vanishing  $\mu$  it exhibits a  peak at the transition
temperature. The height of this peak  depends
not only on the pion mass~\cite{org,orgl} but is also influenced by the value of  the chemical potential. The mesonic fluctuations weaken the
peak structure both at finite and at vanishing quark density. For finite $\mu$
the kurtosis becomes negative following the same trends as  seen in the fourth order cumulant.

\section*{Summary}
We have discussed  the influence of quantum fluctuations near  the chiral phase transition within the Polyakov loop extended
 quark--meson (PQM)  model based on the functional renormalization group method (FRG).  We have shown that  non-perturbative dynamics
introduced in the FRG approach essentially modifies  predictions of the PQM model derived  in the  mean-field approximation. In particular,  we have demonstrated   quantitative changes of the phase diagram
 and modification of  net-quark number fluctuations and their  higher moments.
We have  indicated the role and importance of the ratio  of the fourth to second order  cumulant moments to identify deconfinement and  chiral phase transition in a strongly interacting medium.

%

\end{document}